\documentclass[runningheads]{llncs}
\usepackage[T1]{fontenc}
\usepackage{comment}
\usepackage{amssymb}
\usepackage{amsfonts}
\usepackage{amsmath}
\usepackage{tabularx}
\usepackage{graphicx}
\usepackage{tikz}
\usetikzlibrary{arrows.meta}
\usepackage{xspace}
\usepackage{color}
\usepackage{booktabs}
\usepackage{todonotes}

\usepackage{listings}
\usepackage{upquote}  
\definecolor{imgqlcomment}{gray}{0.45}
\lstdefinelanguage{ImgQL}{%
  morekeywords={let,in,for,do,import,print,save,load,true,false},
  morecomment=[l]{//},
  morestring=[b]",
  sensitive=true,
}
\lstdefinestyle{imgql}{%
  language=ImgQL,
  basicstyle=\ttfamily\scriptsize,
  keywordstyle=\bfseries,
  commentstyle=\color{imgqlcomment},
  stringstyle=\ttfamily,
  numbers=left, numberstyle=\tiny, stepnumber=5, numbersep=6pt,
  breaklines=true, breakindent=0pt, breakautoindent=false,
  postbreak={\mbox{\quad$\hookrightarrow$\space}},
  columns=fullflexible, keepspaces=true,
  frame=tb, framesep=4pt, captionpos=t,
  abovecaptionskip=0.6em, belowcaptionskip=0.7em,
  xleftmargin=1.9em,
  extendedchars=true,
  literate={·}{{\textperiodcentered}}1 {…}{{\ldots}}1
           {‹}{{$\langle$}}1 {›}{{$\rangle$}}1
           {▶}{{$\blacktriangleright$}}1 {◉}{{$\odot$}}1 {⤴}{{$\nearrow$}}1,
}

\AtBeginDocument{}

\usepackage[linesnumbered,ruled]{algorithm2e}

\DontPrintSemicolon
\SetKwFunction{check}{Sat}
\SetKwFunction{checkUntil}{CheckSurr}
\SetKwFunction{checkDiff}{CheckProp}
\SetKwFunction{checkGroup}{CheckGroup}
\SetKwFunction{visit}{Visit}
\SetKwProg{myfun}{Function}{}{}
\SetKwSwitch{Match}{Case}{Other}{Match}{}{case}{otherwise}{~ }{}
\SetKwIF{If}{ElseIf}{Else}{if}{}{else if}{else}{endif}

\SetKw{Let}{let}
\SetKw{For}{for}
\SetKw{Do}{do}
\SetKw{In}{in}
\SetKw{Var}{var}
\SetKw{Conj}{and}
\SetKw{Neg}{!}

\SetKw{Load}{load}
\SetKw{Import}{import}
\SetKw{Save}{save}
\SetKw{Print}{print}

\SetAlFnt{\small}
\SetAlCapFnt{\small}
\SetAlCapNameFnt{\small}
\SetAlCapHSkip{0pt}
\IncMargin{-\parindent}

\newcommand{\RED}[1]{\textcolor{red}{#1}}
\newcommand{\BROWN}[1]{\textcolor{brown}{#1}}

\newcommand{\MAGENTA}[1]{\textcolor{magenta}{#1}}

\newcounter{lbnot} 
\newenvironment{lbnot}[1][]{\refstepcounter{lbnot}\par\medskip
   \noindent \textbf{\RED{NfL~\thelbnot.}  #1} \rmfamily}{\medskip}

\newcounter{mknot} 
\newenvironment{mknot}[1][]{\refstepcounter{mknot}\par\medskip
   \noindent \textbf{\MAGENTA{NfM~\themknot.}  #1} \rmfamily}{\medskip}

\newcounter{vcnot} 
\newenvironment{vcnot}[1][]{\refstepcounter{vcnot}\par\medskip
   \noindent \textbf{\BROWN{NfV~\thevcnot.}  #1} \rmfamily}{\medskip}

\newcommand{\imgql}{\texttt{ImgQL}\xspace}
\newcommand{\slcs}{\texttt{SLCS}\xspace}
\newcommand{\vl}{\texttt{VoxLogicA}\xspace}
\newcommand{\vltwo}{\texttt{VoxLogicA2}\xspace}
\newcommand{\nnunet}{\texttt{nnU-Net}\xspace}
\newcommand{\unet}{\texttt{U-Net}\xspace}

\newcommand{\sitk}{\texttt{SimpleITK}\xspace}

\newcommand{\calA}{\mathcal{A}}

\newcommand{\calC}{\mathcal{C}}
\newcommand{\calD}{\mathcal{D}}

\newcommand{\calI}{\mathcal{I}}

\newcommand{\calM}{\mathcal{M}}
\newcommand{\calN}{\mathcal{N}}

\newcommand{\calV}{\mathcal{V}}

\newcommand{\SET}[1]{\{#1\}}
\newcommand{\ZET}[2]{\SET{#1 \,|\, #2}}

\newcommand{\nats}{\mathbb{N}}

\newcommand{\reals}{\mathbb{R}}

\newcommand{\cs}{CS}
\newcommand{\qdcs}{QdCS}
\newcommand{\closure}{\calC}

\newcommand{\interior}{\calI}

\newcommand{\ap}{\tt AP}
\newcommand{\model}{\calM}
\newcommand{\aeval}{\calA}
\newcommand{\peval}{\calV}

\newcommand{\arel}{R}
\newcommand{\form}{\Phi}
\newcommand{\sem}[1]{[\![#1]\!]}

\newcommand{\lfalse}{{\tt false}}

\newcommand{\ldist}[1]{\calD^{#1}}
\newcommand{\lnear}{\calN}
\newcommand{\lreach}{\rho}
\newcommand{\touch}{{\tt touch}}
\newcommand{\grow}{{\tt grow}}
\newcommand{\smoothen}{{\tt smoothen}}

\newcommand{\perfCases}{40}
\newcommand{\perfColdKernels}{26\,344}
\newcommand{\perfColdWall}{52.5}

\newcommand{\perfCpuPerWallI}{8.0}

\newcommand{\perfCpuPerWallXXIV}{18.5}
\newcommand{\perfFusionLiveOff}{5.1}
\newcommand{\perfFusionLiveOn}{3.8}
\newcommand{\perfFusionOps}{8\,801}
\newcommand{\perfFusionSpeedup}{1.13}

\newcommand{\perfItkOneBestThreads}{16}
\newcommand{\perfItkOneBestWall}{77.3}

\newcommand{\perfItkOneSpeedupVIII}{5.51}
\newcommand{\perfItkOneSpeedupXVI}{6.17}

\newcommand{\perfResumeShare}{57}

\newcommand{\perfSaturationXVI}{0.94}

\newcommand{\perfSpeedupXXIV}{2.19}
\newcommand{\perfStoppedPlusResumedWall}{56.3}

\newcommand{\perfUninterruptedWall}{52.3}
\newcommand{\perfVlAllSpeedup}{2.25}

\newcommand{\perfVlGoals}{840}
\newcommand{\perfVlMaxDiff}{5\cdot 10^{-11}}
\newcommand{\perfVlMemRatio}{1.5}
\newcommand{\perfVlOneWall}{101.7}

\newcommand{\perfVlTwoAllWall}{45.2}
\newcommand{\perfVlTwoWall}{102.7}

\newcommand{\perfWallVIII}{56.9}

\newcommand{\perfWallXXIV}{54.6}

\newcommand{\perfWarmKernelsMax}{1\,099}
\newcommand{\perfWarmWall}{1.39}

\newcommand{\csEvalCases}{51--70}
\newcommand{\csFixedDiceGt}{0.8276}
\newcommand{\csBestDiceGt}{0.8555}
\newcommand{\csBestThrMeanGt}{0.89}
\newcommand{\csBestThrMedianGt}{0.90}
\newcommand{\csModelDice}{0.8588}

\newcommand{\csBestDiceNn}{0.8502}

\begin{document}
%
\title{VoxLogicA2: A Spatial Model Checker for Neuro-symbolic Medical Imaging (extended version)}
%
%
\author{Gina Belmonte\inst{1}\orcidID{0000-0002-7087-8914} \and 
Laura Bussi\inst{2}\orcidID{0000-0003-1292-4086} \and
Vincenzo Ciancia\inst{3}\orcidID{0000-0003-1314-0574} \and
Mieke Massink\inst{3}\orcidID{0000-0001-5089-002X}}
\authorrunning{G. Belmonte et al.}
%
\institute{Azienda USL Toscana Nordovest, Ospedale San Luca, Lucca, Italy\\
\email{gina.belmonte@uslnordovest.toscana.it}  \and
University of Amsterdam, Amsterdam, The Netherlands \\
\email{l.bussi@uva.nl} \and
Istituto di Scienze e Tecnologie dell'Informazione, CNR, Italy
\email{\{vincenzo.ciancia,mieke.massink\}@cnr.it}}
\maketitle              
\begin{abstract}
We present the tool \vltwo, a spatial model checker aimed at Computer Vision, with applications in high-resolution 2D and 3D medical imaging. The new version of the tool combines logical primitives, non-logical primitives, and training and prediction primitives of state-of-the-art neural networks, freely interleaved with the symbolic tool functionality. The primary application field is neuro-symbolic medical image analysis.  \vltwo is a fast, parallelising spatial model checker with a declarative domain-specific language, capable of describing both the learning procedures and the spatial logic procedures of an analysis in one program. We validate our approach by verifying its improved efficiency, and by using the new constructs of the language to describe case studies of previous work, which were based on Python, with a specification that uses \vltwo only.

\end{abstract}
%
%
%

\section{Introduction}

Medical image analysis requires perception abilities as well as symbolic reasoning and knowledge representation capabilities. When analysing an image, we may be interested in enforcing some constraints, or exploit domain specific knowledge. Today this analysis is most often carried out by neural networks, which perceive at the level of a human expert but do not reason symbolically. The BraTS challenge~\cite{Menze2015} is a competition among different methods performing high grade glioblastoma segmentation. Machine Learning (ML) methods achieve very high accuracy, which is crucial as segmentation is the basis for follow up therapies, like radiotherapy. Still, the predictions are not explainable nor auditable, which are desired properties for a physician to understand the procedure leading to a certain result. It is thus natural to ask whether it is possible to combine these two approaches in order to get efficient, accurate, auditable, reproducible and explainable results.


The field that combines symbolic and neural methods, neuro-symbolic AI, is a particularly active area of research (see e.g.~\cite{DeRaedt2026,Choudhary2026,MJL2023}). Its application to image analysis is motivated by the complementary strengths and weaknesses of neural and symbolic representations: neural models are effective at extracting semantic information from visual data, but are weak in providing explanations for their results. Symbolic representations, instead, can explicitly capture domain knowledge and relations between the spatial entities identified in an image, but are weaker in capturing the variability of real-world datasets. A careful combination of the two approaches may give the advantages of both and increase the reliability and the explainability of the image analysis. This requires however a suitable platform to operate such combined methods that is transparent, auditable and permits easy and controlled replication of the results. This is the main aim of \vltwo.

In recent years, \emph{spatial logic} emerged as a suitable symbolic language to express properties (such as reachability, distance, similarity, shape and proximity) of entities located in logical or physical spaces, among which general graphs, polyhedra and digital images. Ciancia et al.~\cite{CLLM14,CLLM16,BCLM19} introduced the Spatial Logic for Closure Spaces (\slcs), a spatial logic interpreted over \emph{C\^{e}ch closure spaces}, a generalisation of topological spaces. Digital images are symmetric regular graphs, whose nodes are pixels (voxels in 3D) and whose edges are given by adjacency, so \slcs can state properties of images.

Along with \slcs\ various spatial model checkers have been developed, among which \vl~\cite{BCLM19,CianciaBLM21}, which is specialised and highly optimised for digital images. \vl obtained results on par with the state of the art on the Brain Tumour Segmentation challenge (BraTS)~\cite{Menze2015} dataset, both in terms of accuracy, compared to provided expert ground truth segmentation, and in terms of performance, i.e. computational resources and time needed to obtain the results. See~\cite{CianciaLM25,BelmonteCLM26} for an overview on recent work on spatial model checking and~\cite{CianciaBLM21} for a hands-on tutorial for \vl.

In this paper, we present  \vltwo, a new version of \vl 
that includes AI primitives for training and prediction, providing an innovative platform for neuro-symbolic image analysis. \vltwo integrates \nnunet~\cite{Isensee2021}, a state-of-the-art, self-configuring deep learning framework for biomedical image segmentation. Combined with symbolic spatial model checking procedures, \vltwo can perform hybrid analysis based on a fully certified process in the sense that all steps in the procedure, from the configuration of the learning process to the symbolic specification and production of results are transparent, fully auditable and independently replicable. This represents a step towards the EU Artificial Intelligence Act~\cite{EUAIAct24} and the EU Ethics Guidelines for Trustworthy AI~\cite{EUAIguide2019}, which both ask for trustworthy and explainable AI procedures. 

\vltwo is a new implementation of \vl, written in Python, the language of the neural network ecosystem, so that a hybrid analysis can call a network directly. A program is turned into a Merkle directed acyclic graph (DAG)~\cite{Mer88,Benet14}, in which identical sub-expressions become one node~\cite{FC06} whose identity is the hash of its contents, so that a sub-expression is computed once, however often it occurs. The same hashes key a persistent cache: a value computed by one run is available to every later run, and an interrupted analysis is resumed rather than restarted. Execution is parallel and demand driven, with coordination on a single lock-free event loop, while the imaging kernels fill the cores of a workstation. The design targets coarse-grained work, where the cost of scheduling is negligible compared to that of computing, which is the regime in which automatic parallelisation pays.

We illustrate \vltwo on a realistic case study, a threshold sweep in brain tumour segmentation and a form of neuro-symbolic analysis, both proposed in~\cite{BCM25} using separate tools in a non-integrated way.

\begin{figure}[tb]
\centering
\includegraphics[width=0.31\textwidth]{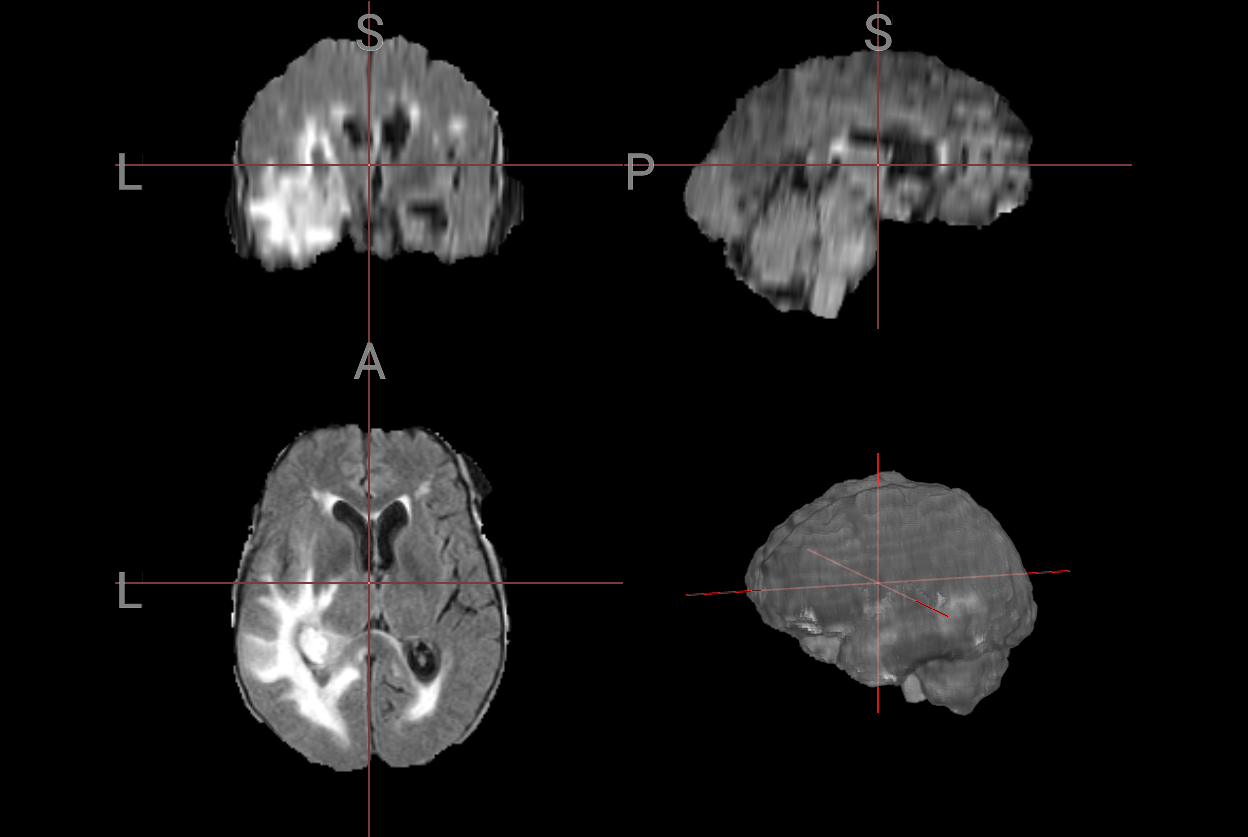}\;\;
\includegraphics[width=0.31\textwidth]{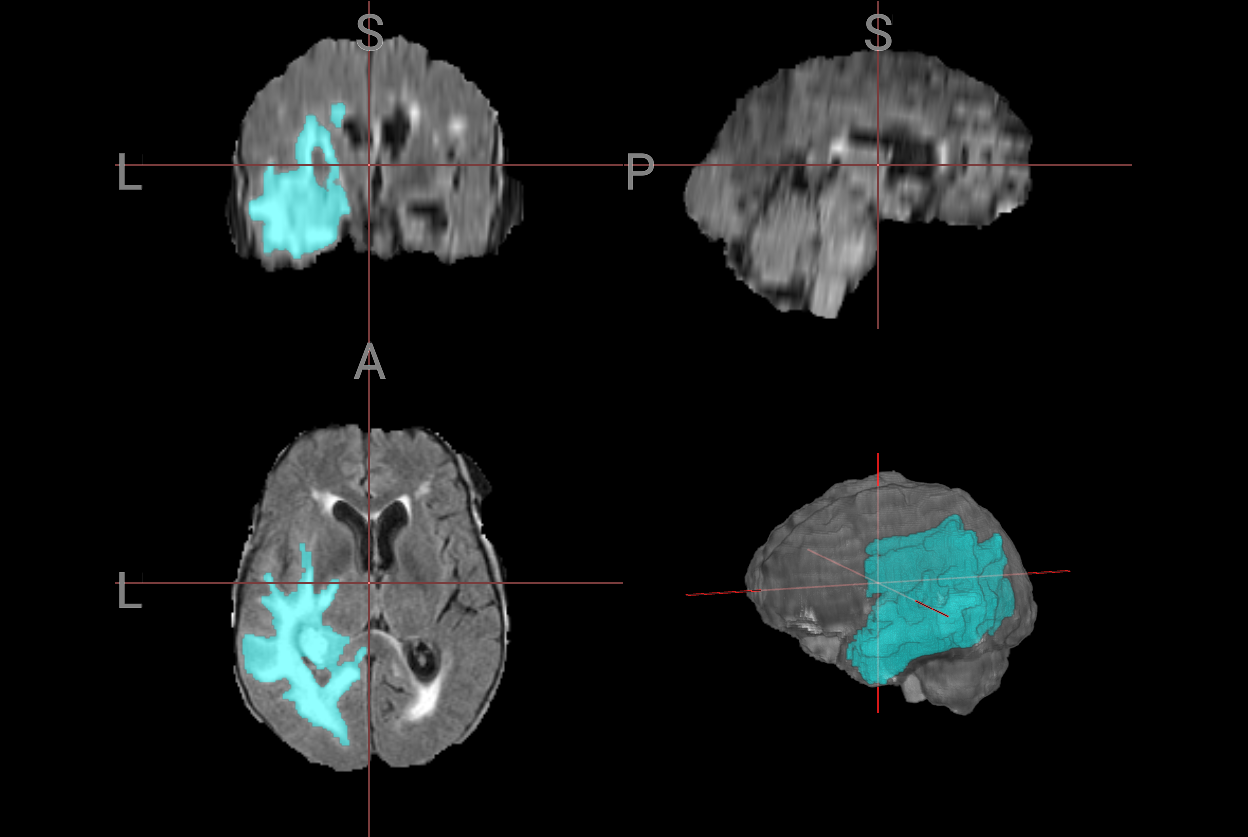}\;\;
\includegraphics[width=0.31\textwidth]{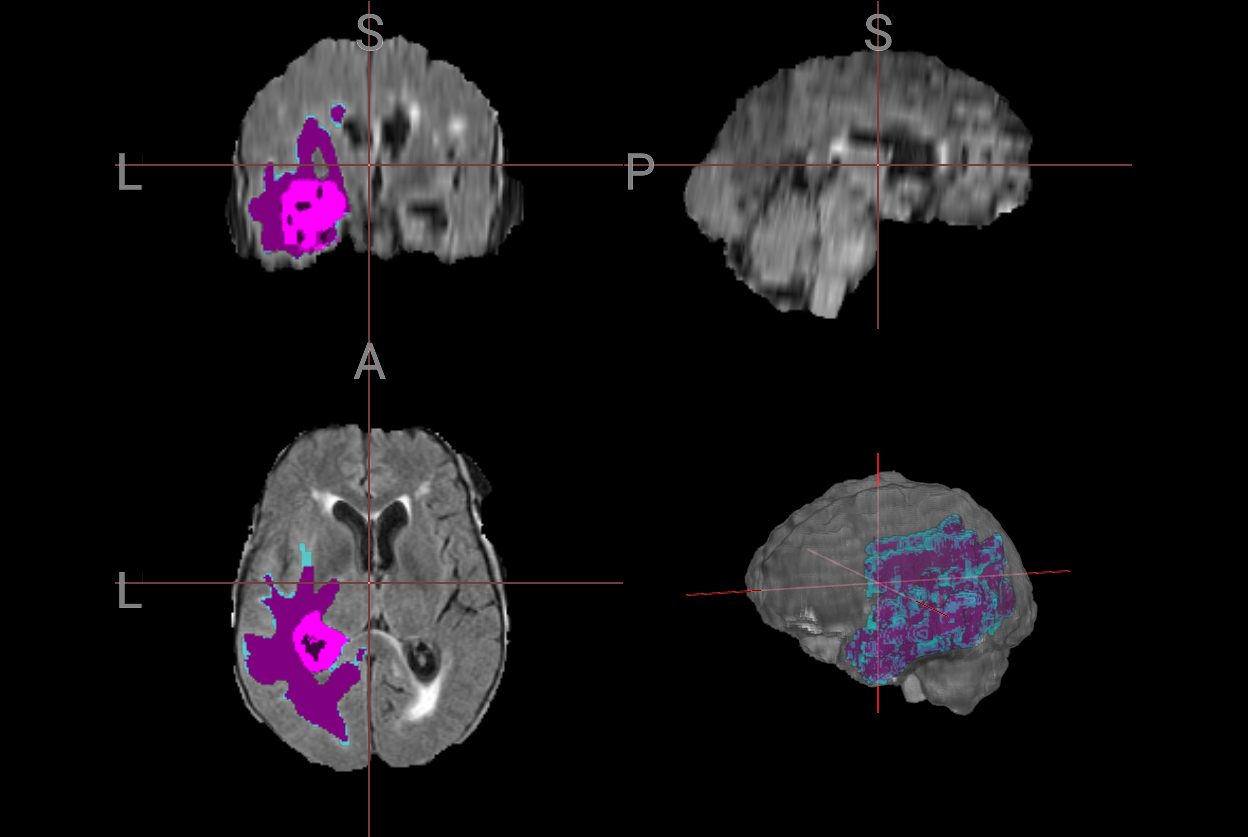}\;\;
\caption{Example brain tumour segmentation of BraTS 2020 case BraTS20\_Training\_165 (TCIA01\_147 in BraTS 2019). From left to right: Original MRI image; Symbolic tumour segmentation with \vl in cyan; Expert segmentation (ground truth) overlay in magenta. Images produced with \vl-UI.}
\label{img:segmentation}
\end{figure}

\paragraph{Case study 1: brain tumour segmentation.} The first case study is the segmentation of brain tumours on the BraTS2020 dataset, measured against expert segmentations (Fig.~\ref{img:segmentation}). The procedure is that of~\cite{BCLM19}, which obtained results on par with state-of-the-art machine learning tools on the 2017 edition of BraTS. We run both \vl\ and \vltwo\ on it to compare their results and performance.

\paragraph{Case study 2: threshold sweep and neuro-symbolic analysis.} The segmentation depends on two thresholds: {\tt hiThr} selects the \emph{hyper-intense} voxels, and a fixed value always finds the tumour core; {\tt viThr} selects the very intense voxels of the surrounding oedema, and can be adjusted per case to gain accuracy. We sweep 21 values of {\tt viThr} in two experiments: first against the expert ground truth, then, in the neuro-symbolic one, against the prediction of a trained \nnunet\ model, used as a pseudo-ground-truth against which {\tt viThr} is calibrated.

\paragraph{Synopsis.} In Section~\ref{sec:back} we recall basic notions about \slcs, \vl and \nnunet. Section~\ref{tool} presents the architecture of \vltwo and how it was developed and tested. Section~\ref{sec:test} details the case studies and Section~\ref{sec:perf} the performance analysis; Section~\ref{sec:conc} concludes the paper including an outlook to future research.


\section{Background}\label{sec:back}

\subsection{Spatial logic and spatial model checking with \vl}
We first briefly recall relevant notions on closure spaces, a generalisation of topological spaces, on which our spatial logic and our spatial model checking approach are based. Closure spaces include interesting structures, such as general graphs. We refer
to~\cite{Gal99,Gal03,RLG13,Gal14} and to our previous work~\cite{BCLM19,Ba+20,CLLM14,CLLM16} for further details on the relevant theoretical aspects.

\begin{definition}[Closure Space]
A \emph{closure space} ---\cs\ for short--- is a pair $(X,\closure)$ where $X$ is a set of points and the closure operator $\closure : 2^X \to 2^X$ is a function that satisfies the following three axioms:
\begin{itemize}
	\item $\closure(\emptyset) = \emptyset$
	\item $A \subseteq \closure(A)$ for all $A \subseteq X$
	\item $\closure(A_1 \cup A_2) = \closure(A_1) \cup \closure(A_2)$ for all $A_1, A_2 \subseteq X$
\end{itemize} 
\end{definition}

Closure spaces are a generalisation of topological spaces in the sense that the closure operator is characterised by only three of the usual four axioms from topology, omitting the fourth one that requires that the closure operator is idempotent.
The {\em interior} of a set $Y\subseteq X$ is obtained by duality, i.e. $\interior(Y)=\overline{\closure(\overline{Y})}$
where $\overline{Y}=X\setminus Y$ is the complement of $Y$.

The logic underlying the declarative language \imgql is the Spatial Logic for Closure Spaces (\slcs). Its syntax is given by the syntax of propositional logic, extended with a modal reachability operator.

\begin{definition}[Spatial Logic for Closure Spaces]
Given a set $AP$ of atomic propositions, the grammar generating the language \slcs is defined as follows:
\[ \Phi := a \mid \top \mid \lnot \Phi \mid \Phi \land \Phi \mid \lreach \ \Phi [\Phi] \mid \ldist{I} \Phi \ \mbox{for} \ a \in AP \]
\end{definition}

The operators $\lnot$ and $\land$ are the usual Boolean operators of negation and conjunction, respectively. The spatial reachability operator $\lreach \ \Phi [\Phi]$ is defined on paths through space defined as follows. Let $(\nats,\closure_{succ})$ be the \cs\ of  the natural numbers $\nats$ with  the binary successor relation $succ = \ZET{(m,n) \in \nats^2}{n=m+1}$. A (discrete) {\em path} over \qdcs\ $(X,\closure)$ is a continuous function\footnote{A {\em continuous} function from \cs\ $(X_1,\closure_1)$ to \cs\ $(X_2,\closure_2)$ is a function $f: X_1 \to X_2$ such that $f(\closure_1(Y)) \subseteq \closure_2(f(Y))$ for all $Y \subseteq X_1$.}  from $(\nats,\closure_{succ})$ to $(X,\closure)$.
A point $x$ in space satisfies the formula $\lreach \ \Phi_1 [\Phi_2]$ if it can reach a point that satisfies $\Phi_1$ along a path consisting of points all satisfying $\Phi_2$. The distance operator $\ldist{I}$ is defined as follows. A point $x$ satisfies $\ldist{I} \Phi$ if the distance between $x$ and the points satisfying $\Phi$ is within interval $I$, where interval $I\subseteq \reals_{\geq 0}$ of non-negative real numbers. The distance function $d: X \times X \to \reals_{\geq 0} \cup \SET{\infty}$ satisfies $d(x,y)=0$ if and only if $x=y$; $d$ is lifted to sets in the usual way: $d(x,\emptyset)=\infty$ and for $\emptyset \subset Y \subseteq X$ we have $d(x,Y)=\inf\ZET{d(x,y)}{y \in Y}$.

Given any relation $\arel \subseteq X \times X$, $(X,\closure_{\arel})$, with $\closure_{\arel}(Y) = Y\cup \ZET{x}{\exists y \in Y. y \, \arel \,x}$,  is a \cs{,} specifically, a quasi-discrete \cs\ (\qdcs). In particular, a digital image can be modelled as a finite \cs\ (every finite \cs\ is also a \qdcs{)} where $X$ is the set of voxels and $\arel$ their {\em adjacency relation}\footnote{All the theory and related model checkers work both for 2D and 3D even though we use only 3D in the present paper. In the current work we use the word `voxel' for 3D `pixels'.}. In the present paper, we will consider only \qdcs{s} and use the {\em ortho-diagonal} relation in 3D, i.e. the reflexive and symmetric relation where two voxels are {\em adjacent} if and only if they share a face, an edge or a vertex. Other presentations of \slcs\ (see e.g.~\cite{BCM25}) include \emph{forward} and \emph{backward} reachability operators. In this paper, though, we deal with images, represented as symmetric adjacency graphs, and the two operators turn out to be equivalent. Hence the choice of including here just one, the forward reachability operator. 

\slcs{} is interpreted over {\em attributed distance closure models}, i.e. structures $((X,\closure),d,\aeval, \peval)$ where $(X,\closure)$ is a \cs{,} $d$ and $\aeval$ are distance and attribute evaluation functions, respectively, and $\peval: \ap \to 2^X$ is a valuation which maps the  {\em atomic predicates} to the points satisfying them.
The value $\aeval(a,x) \in V$ of attribute $a$ of point $x$ is given by the attribute evaluation function
$\aeval: A \times X \to V$, for suitable set $V$ of {\em values}.

The satisfaction relation for  model
$\model = ((X,\closure),d ,\aeval,\peval)$, $x \in X$, and  formulas $\form$ is defined recursively on the structure of \slcs formulas $\form$ as follows, where
$\sem{\form}{}^{\model} = \ZET{x\in X}{\model,x \models \form}$ is the set of points in $\model$ satisfying $\form$:



\begin{definition}
{\em Satisfaction} $\model, x \models \phi$  of an \slcs formula $\phi$ at point $x \in X$ is defined as follows, where
$\sem{\form}{}^{\model} = \ZET{x\in X}{\model,x \models \form}$ is the set of points in $\model$ satisfying $\form$.
\[
\begin{array}{r c l c l c l l}
\model,x & \models  & p & \Leftrightarrow & x  \in V(p)\\
\model,x & \models  & \lnot \,\phi & \Leftrightarrow & \model,x  \not \models \phi\\
\model,x & \models  & \phi_1\, \land \,\phi_2 & \Leftrightarrow &
\model,x  \models \phi_1 \mbox{ and } \model,x  \models \phi_2\\
\model,x & \models  & \lreach \;\phi_1[\phi_2] & \Leftrightarrow & \mbox{path } \pi 
\mbox{ and index } k 
\mbox{ exist s.t. }
\pi(0)=x \mbox{ and } 
\model, \pi(k) \models  \phi_1 
\\
& &  & & 
\mbox{and for all indexes } j: 0 < j  < k \mbox{ implies } \model, \pi(j) \models \phi_2\\ 
\model,x & \models & \ldist{I} \form              & \Leftrightarrow & d(x,\sem{\form}{}^{\model}) \in I.
\end{array}
\] 
\end{definition}

{\emergencystretch=2em The brain tumour segmentation uses the derived operators
$\lnear \form \equiv \lreach \, \form \,[\lfalse]$,
$\touch(\form_1 , \form_2) \equiv \form_1 \land \lreach\, \form_2[\form_1]$,
$\grow(\form_1 , \form_2) \equiv \form_1 \lor \touch( \form_2 , \form_1)$ and
$\smoothen(\form, r) \equiv \ldist{\leq r} (\ldist{\geq r} \lnot\form)$.
Point $x$ satisfies $\lnear \form$ if it satisfies $\form$ or it is adjacent to a point satisfying $\form$. Point $x$ satisfies $\touch(\form_1 , \form_2)$ if it satisfies $\form_1$ and from it one can reach a point satisfying $\form_2$ only passing by a sequence of adjacent points satisfying $\form_1$.\par}
The grow operator $\grow(\form_1 , \form_2)$ extends the set of points satisfying $\form_1$  with those points satisfying $\form_2$  that can reach points satisfying $\form_1$ through paths the points of which satisfy $\form_2$. In other words, the area satisfying $\form_1$ is extended with an area of points satisfying $\form_2$ under the condition that these areas touch each other. The smoothen operator first erodes and then dilates the area of points satisfying $\form$ by an amount of $r$, obtaining a smoothening effect.


\subsection{AI for imaging and \nnunet}
Biomedical image analysis increasingly relies on deep neural networks, and in particular on \unet~\cite{RFB2015}, the most widely adopted architecture for medical image segmentation, to automatically extract meaningful representations from medical images. Among these tasks, semantic segmentation aims at assigning a semantic label to each pixel or voxel of such images, enabling the delineation of anatomical structures or pathological regions of interest.

Despite their accuracy, the deployment of such networks in actual medical workflows is often hindered by the expertise required to configure them and by the black-box nature of the resulting learned model. Achieving high quality segmentation typically requires substantial dataset-specific choices concerning preprocessing, network configuration, training, and post-processing. \nnunet~\cite{Isensee2021} addresses the first problem with a self-configuring method for medical image segmentation, a strong baseline that often matches or outperforms highly specialised methods. The second problem, the lack of explainability and transparency, cannot be solved by configuration; here the symbolic approach plays a complementary role, providing a domain-oriented, logic-based description of the segmentation.
\section{Tool Architecture and Implementation}\label{tool}

The implementation of the precursor of \vltwo{,} \vl{,} is written in F\# and is itself parallelised on CPU; an on-GPU version, used for video-stream analysis, was introduced in~\cite{BussiCG21}, as well as a temporal extension~\cite{BCGM26} for the analysis of spatio-temporal properties, according to the language introduced in~\cite{BussiCG23}. \vltwo is implemented entirely in Python, as it is the preferred language in the area of neural networks. Although Python is interpreted, \vltwo matches the speed of \vl\ on a single case and is faster on a whole dataset (Section~\ref{sec:perf}), thanks to native libraries, the \emph{Numba} just-in-time compiler~\cite{LamPS15}, and the release of the \emph{global interpreter lock} in modern Python.

\subsection{\vltwo\ Architecture}
\begin{figure}[tb]
\centering
\resizebox{0.85\textwidth}{!}{%
\begin{tikzpicture}[
  font=\footnotesize,
  box/.style={draw, rounded corners=2pt, align=center, inner sep=3pt,
              minimum height=9mm, minimum width=20mm},
  arr/.style={-{Stealth[length=1.7mm]}, semithick},
  lab/.style={font=\scriptsize, midway}]
  \node[box] (src)   at (0, 0)    {Program\\{\scriptsize\texttt{.imgql} file}};
  \node[box] (parse) at (3.1, 0)  {Parser};
  \node[box] (red)   at (6.2, 0)  {Reducer\\{\scriptsize shared sub-expressions}};
  \node[box] (plan)  at (9.6, 0)  {Symbolic plan\\{\scriptsize Merkle DAG}};
  \node[box] (tc)    at (12.8, 0) {Type checker};
  \draw[arr] (src) -- (parse);
  \draw[arr] (parse) -- node[lab, above] {AST} (red);
  \draw[arr] (red) -- (plan);
  \draw[arr] (plan) -- (tc);
  \node[box] (exp)   at (1.4, -2.6)  {Expander\\{\scriptsize unrolls loops}};
  \node[box] (sched) at (6.8, -2.6)  {Scheduler\\{\scriptsize dependency graph}};
  \node[box] (exec)  at (12.2, -2.6) {Executor\\{\scriptsize thread pool}};
  \draw[arr] (plan.south) -- node[lab, right] {goals} (plan.south |- sched.north);
  \draw[arr] ([yshift=1.5mm]sched.east) -- node[lab, above] {ready nodes} ([yshift=1.5mm]exec.west);
  \draw[arr] ([yshift=-1.5mm]exec.west) -- node[lab, below] {results} ([yshift=-1.5mm]sched.east);
  \draw[arr] ([yshift=1.5mm]sched.west) -- node[lab, above] {sequences} ([yshift=1.5mm]exp.east);
  \draw[arr] ([yshift=-1.5mm]exp.east) -- node[lab, below] {new nodes} ([yshift=-1.5mm]sched.west);
  \node[box, minimum width=62mm] (store) at (9.4, -4.9)
        {Value store\\{\scriptsize memory cache and disk cache (SQLite)}};
  \node[box] (out) at (1.4, -4.9) {Output\\{\scriptsize\texttt{print}, \texttt{save}}};
  \draw[arr, {Stealth[length=1.7mm]}-{Stealth[length=1.7mm]}]
        (exec.south) -- node[lab, right] {values} (exec.south |- store.north);
  \draw[arr] (store.west) -- (out.east);
\end{tikzpicture}}
\caption{The \vltwo\ execution pipeline.}
\label{img:pipeline}
\end{figure}
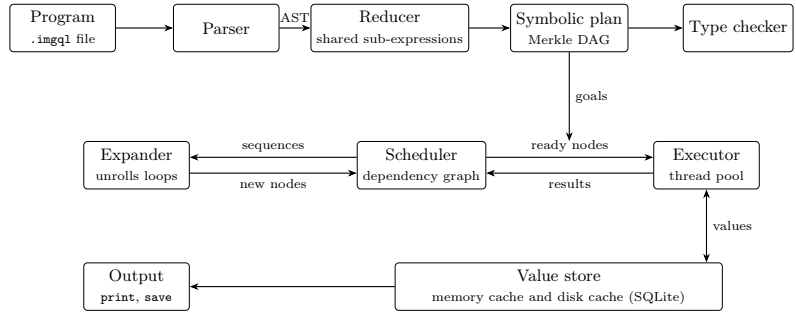

At its heart, \vltwo is a general-purpose parallelising computation engine for problems, such as imaging, where the execution time of computation nodes vastly dominates that of scheduling. Its core is a content-addressed Merkle DAG of expressions, whose results are memoised in a memory and a disk cache. The execution pipeline is shown in Fig.~\ref{img:pipeline}.

The architecture has two macro-components: the front-end, composed of components summarised in Table~\ref{tab:frontend}, and the execution engine. Both are described below.

\subsubsection{Front-end of \vltwo.}

\begin{table}[tb]
\centering\small
\caption{Front-end: from source text to an executable plan.}
\label{tab:frontend}
\begin{tabularx}{\textwidth}{@{}l X@{}}
\toprule
\textbf{Stage} & \textbf{Role} \\
\midrule
Parser             & Text to AST; Lark grammar, syntax only, no name resolution. \\
Reducer            & AST to symbolic plan; resolves bindings and imports, lowers constructs to primitive calls, hash-conses identical sub-expressions into one node. \\
Symbolic plan      & Immutable Merkle DAG; a node's id is the hash of its operator, of the ids of its children, of its attributes and of the kind of value it produces, independent of construction order. \\
Type checker       & Abstract execution of the plan with types in place of values; each primitive supplies its own type rule. \\
Primitive registry & One spec per operator: symbolic planner, runtime kernel, type rule, and the flags the engine reads (elementwise, lazy, rewrite). \\
\bottomrule
\end{tabularx}
\end{table}

\vltwo takes as input a declarative specification in the Image Query Language (\imgql). The module \texttt{Parser} parses this into an Abstract Syntax Tree (AST). The \texttt{Reducer} turns the AST into the \texttt{SymbolicPlan}, an append-only Merkle DAG~\cite{Mer88} in which two identical sub-expressions have the same identity and are stored once, whatever the order in which the graph was built. Nothing is executed at this stage: the plan only records what the engine will compute.

The declarative language \imgql\ includes all the \slcs operators, a number of domain specific derived operators, and operators for manipulating images. For \vltwo the language has been extended, since Python gives a faster path to many libraries that are useful in the application domain. A wide range of imaging primitives, based on the Insight Tool Kit (ITK), an open source, cross platform library for (medical) image analysis, through \sitk~\cite{Yaniv2018}, now covers the whole spectrum of image analysis, and all the primitives of \vl are available as well. The language also has primitives for managing collections of data. Among them, \texttt{map} applies a function to every element of a sequence, and \texttt{fold} combines the elements of a sequence into a single value, using an operator such as addition or maximum.

\subsubsection{Execution engine.} 
Once the \texttt{SymbolicPlan} has passed type checking, the engine executes it on demand: only the operations that contribute to a goal (\texttt{print} or \texttt{save}) are computed. All coordination runs on a single event loop; only the kernels, the writes to disk and the unrolling of loops run on other threads. The engine consists of the following components.

\paragraph{Node table and store.} The node table records, for each node of the plan, its state and, once computed, its value. Since the identity of a node is a hash of its contents, it is the same across threads and processes, and it serves as the key of a two-level cache: values in use are kept in memory, and every computed value is also written to a store on disk together with its \emph{recipe}, that is, the operator, the identities of the arguments and the attributes (such as the value of a constant). Recipes and small values are kept in an SQLite database, while large values such as images are written as separate files next to it, outside the database. A cache hit is therefore not limited to the current session: running the same program twice returns immediately on the second run, and so does a \emph{different} program that shares sub-expressions with the first one. A node is dispatched at most once. The cache can be disabled via program options.

\paragraph{Dependency graph.} A node becomes ready to run when the last of its inputs has been computed, as in Kahn's topological sorting~\cite{Kahn62} applied incrementally: when a node completes, only the nodes that consume it are updated, so the cost of scheduling depends on the frontier of the computation and not on the size of the plan. The graph also counts, for each value, the consumers that still have to run; a value is dropped from memory as soon as its last consumer has run, so the working set follows the frontier of the computation. Dropping a value is always safe, because readiness depends on \emph{completion} and not on the value being in memory: an evicted value costs at most a reload from the store, or a recomputation when the store is disabled.

\paragraph{Ready queue.} Nodes that are ready wait in a priority queue. A node inherits its priority from the queries that demand it, so that, for instance, an item of a sequence that the user asked for is computed before the rest. When memory is scarce, the queue holds back ready nodes rather than starting them, but never to the point of leaving all workers idle.

\paragraph{Executor.} Kernels run on a pool of worker threads. Each kernel is a pure function of its inputs, and the libraries used (ITK through \sitk, NumPy, Numba) release the Python global interpreter lock while they compute, so the workers run in parallel. Since ITK filters are themselves multi-threaded, parallelism between operations and within an operation combine.

\paragraph{Fusion.} Many operations of an image analysis are element-wise: the value of each voxel of the result depends only on the same voxel of the inputs, as for thresholds, Boolean connectives and arithmetic. When a worker takes such an operation from the ready queue, the engine groups it with the element-wise operations that consume its result and whose other inputs are already available, up to a fixed size. The group, called a \emph{cone}, never includes ITK filters such as distance transforms or morphology, which end it. The cone is dispatched as a single task and, when possible, compiled with Numba into a single loop over the voxels, so the intermediate images inside the cone are never stored. Section~\ref{sec:perf} measures the effect of fusion.

\paragraph{Expander.} Loops whose sequence is only known at run time are unrolled during execution. Unrolling is done in batches on a separate thread, and new loop bodies are started only when the workers would otherwise run out of work, so that a loop over thousands of elements does not flood the scheduler or the memory.

\paragraph{Persister.} Computed values are written to disk by a separate thread, so that writing never stalls computation. If the disk falls behind, the engine slows down the dispatch of new nodes instead of letting unwritten values accumulate in memory.

\paragraph{Memory governor.} The governor bounds the resident memory of the process as measured by the operating system, rather than the engine's own count of the values it holds, which misses, for example, the internal buffers of the imaging libraries.

Primitives such as \texttt{map} and \texttt{fold} are not easy to implement in a semantics based on nodes such as ours: the identity of a node is fixed by the expression it computes, but the elements of a sequence, and often their number, are only known once the sequence has been computed, so the nodes that a \texttt{map} stands for cannot all be created before execution starts.

Iteration has a single semantics, which is reduction. The body of a \texttt{for} or a \texttt{map} is never interpreted to a value: as soon as the iterable is computed, the body is reduced once per element into new nodes, exactly as the reducer does at compile time for a constant sequence, and the resulting nodes are hash-compared against the ones already known, so whatever the elements share is computed once. Iteration therefore re-enters the same scheduling cycle as every other node, and the DAG remains the only place where the meaning of a program lives.

An interrupted computation can be resumed from the cache. The computation
boundary (the nodes that are ready but not yet computed) is written to the
store at a fixed interval, and when the body of a \texttt{for} has been reduced over every element of a sequence,
the fact that the reduction is complete is recorded together with the nodes it
produced, so that a later run is not obliged to reduce the body again merely to
learn the names of nodes whose values it already holds.

Against starvation, the engine checks during the run that every waiting node
is ready, or waits for a value that is being computed, is queued, or belongs to
an iteration still being unfolded; otherwise it fails.

\subsubsection{Type Checking.}
Type checking is an abstract interpretation of the plan: every operator has a type-only counterpart, a function from the types of the arguments to the type of the result. Types are primitive (\texttt{int}, \texttt{float}, \dots) or structured (\texttt{sequence}, \texttt{map}, \texttt{record}).

The arithmetic, comparison and boolean operators support overloading. These
operators all share the same shape: if either operand is an image, the result
is an image; otherwise both operands are scalars, where \texttt{bool} counts as
a scalar beside \texttt{number}, since the scalar paths coerce whatever they
are given. The same operators also broadcast over sequences: whenever one
operand is a sequence, the operation is applied element by element, so the type
of \texttt{xs + 1} is the type of \texttt{x + 1} wrapped in a sequence, and
recursively so for a sequence of sequences. Types are related by a consistent
subtyping relation. Ignoring \texttt{any}, it is a partial order: the numeric
types form a chain, \texttt{int} $\leq$ \texttt{float} $\leq$ \texttt{number},
and structured types are ordered componentwise according to their variance. The
top element \texttt{any} is deliberately compatible with every type in both
directions, which makes the relation non-transitive. Where a single static type must describe several runtime
values, such as the elements of a sequence, or the accumulator of a fold, the
checker computes an upper bound of their types, widening to \texttt{any} when
they share no common supertype.

There are no type variables and no unification: polymorphism lives in the
type rules, which are arbitrary functions of the argument types, so type
checking works bottom-up in a single pass. A rule is written in the
specification of a primitive, built with combinators, or inferred from the
annotations of the Python kernel.

\subsection{Implementation}

The implementation of the tool was supported by the use of agentic AI for
optimisation and debugging. We primarily used the Claude Opus 5 model, and
the generated code was tested with unit tests (via \texttt{pytest}) and against \vl via differential testing
(\texttt{tests/differential}). The approach was human-centric: rather than
asking the model to generate code from scratch, we provided a reference
implementation written by hand. It can still be invoked, through a command line
option, although it does not reach the efficiency and the scalability of the
computation engine that the model produced. The \texttt{imgql} programs used
in the experiments are also human-authored (from previous publications of the
working group). The employed LLM was then used to create a highly efficient
execution engine, which is responsible for parallel execution,
scheduling, memory management, two-level caching, loop expansion and kernel
fusion. Testing AI-generated code with AI-generated tests risks circularity, so
our tests are independent of the generated code in four ways:

\begin{itemize}
    \item differential testing against \vl on the same programs;
    \item the engine is validated against the (hand-written) lazy strategy on
          real data;
    \item the pipelines published in~\cite{BCLM19,BCM25} are re-run, and the
          qualitative findings they report are confirmed; the two artifact
          experiments that do not train a network {\em reproduce bit-identically}
          across runs and across machines;
    \item the engine fuses chains of elementary operations into single
          kernels; property tests check that a fused kernel produces
          exactly the same output, bit for bit, as the sequence of operations
          it replaces.
\end{itemize}

\subsection{Differential Testing}
\vl~\cite{BCLM19} is the first iteration of the same tool, and was written in F\#. The two tools do not implement exactly the same language: the new version is a superset. On the fragment they share, we use
\vl\ as an \emph{oracle} for the corresponding \vltwo primitives. Three evaluators have been run on the same programs, and their results
were compared numerically at a tolerance of $10^{-9}$: the \vl\ binary, and
two configurations of \vltwo, the lazy strategy and the scheduling
engine. The three evaluators agreed on every test program.


\section{Case study}\label{sec:test}

In this section, we illustrate a case study consisting of a neuro-symbolic approach to brain tumour segmentation, developed step by step.
All experiments use the public BraTS 2020 training set~\cite{Menze2015,Bak+17,BakasReyes18}: 293 cases of high grade glioma (HGG) and 76 of low grade glioma, each consisting of pre-operative MRI scans in four modalities (T1, contrast-enhanced T1, T2 and FLAIR), stored as NIfTI files\footnote{\url{https://nifti.nimh.nih.gov}}. The symbolic method uses only FLAIR; the neural model uses all four modalities. The first 50 HGG cases are used to train the network and the following 20 for evaluation. As usual in the BraTS challenge, a segmentation is scored by its Dice similarity index against a reference segmentation.

The symbolic procedure (Section~\ref{SymGTV}) depends on two thresholds, for hyper-intense and for very intense voxels, expressed as percentiles of the intensity of the brain. The first can be fixed reliably, but choosing the second case by case improves accuracy considerably~\cite{BCM25}; Section~\ref{OptSymGTV} specifies the corresponding threshold sweep. Section~\ref{neuroSym} trains \nnunet\ and uses its prediction as a surrogate reference to choose the threshold automatically, without expert annotation, as proposed in~\cite{BCM25}. If the symbolic segmentation then agrees closely with the network, two independent methods find (almost) the same tumour, and the symbolic one comes with an explanation that a domain expert can inspect.

\subsection{Symbolic Whole Tumour Segmentation and Threshold Sweep}\label{SymGTV}\label{OptSymGTV}
The derived operators of Section~\ref{sec:back} ({\tt grow}, {\tt smoothen}, {\tt touch}) are built into \imgql, as well as {\tt leq\_sv} and {\tt geq\_sv} (comparison of an image with a scalar), {\tt percentiles} (percentile rank of the voxels of a given area), {\tt border} (the voxels at the border of the image) and {\tt volume} (the number of voxels of an area). Program~\ref{lst:sym} shows the segmentation procedure for the whole tumour (i.e. the tumour core and adjacent oedema) and the procedure to perform a per-case threshold sweep for the intensity threshold for very intense voxels ({\tt vi\_thr}).

\begin{lstlisting}[style=imgql,float=tb,numbers=none,label={lst:sym},caption={Symbolic segmentation and per-case threshold sweep (excerpt).}]
hi_thr = 0.93
preprocess(flair) =
  let background = touch(leq_sv(0.1, flair), border(flair)) in
  percentiles(flair, not(background), 0)
segment(pflair, vi_thr) =
  let hyper = smoothen(geq_sv(hi_thr, pflair), 5.0) in
  let very  = smoothen(geq_sv(vi_thr, pflair), 2.0) in
  grow(hyper, very)
whole_tumour(gt_image) = geq_sv(1.0, gt_image)

dice(a, b) = 2 * volume(and(a, b)) / (volume(a) + volume(b))

pflair_of(i)    = preprocess(intensity(ReadImage(index(flair_paths, i))))
reference_of(i) = whole_tumour(intensity(ReadImage(index(gt_paths, i))))
sweep_of(i) =
  let pflair = pflair_of(i) in
  let truth  = reference_of(i) in
  for vi_thr in vi_thresholds do dice(segment(pflair, vi_thr), truth)
best_dice   = for i in cases do let s = sweep_of(i) in index(s, argmax(s))
best_vi_thr = for i in cases do index(vi_thresholds, argmax(sweep_of(i)))
\end{lstlisting}

The function {\tt preprocess} normalises the FLAIR image. Images acquired with different scanners and protocols may vary in absolute intensity, so thresholds are applied to percentiles rather than to intensities. The background consists of the voxels from which the border of the image can be reached only passing  through very low intensity voxels; percentiles are computed over the remaining voxels, i.e.\ those being part of the brain. The function {\tt segment} computes the hyper-intense voxels (above {\tt hi\_thr}, the 7\% most intense voxels) and the very intense ones (those above the {\tt vi\_thr} threshold), smoothens them with radius 5 and 2\,mm respectively, and grows the former, high intense area, into the touching parts of the latter, very intense, areas. The reference  ground truth ({\tt whole\_tumour}) image consists of the voxels with a tumour label, i.e.\ of value 1 or more in the ground truth image that was segmented manually by domain experts. The function {\tt dice} shows how the similarity index is calculated between a segmentation {\tt a} and a reference segmentation {\tt b}. It makes use of the function {\tt volume} that counts the number of voxels in a given area. The function {\tt and} of two segmentations {\tt a} and {\tt b} gives the voxels where {\tt a} and {\tt b} overlap.

For each case {\tt i}, {\tt sweep\_of} computes the Dice index of the segmentation against the expert annotation for each of the 21 thresholds from 0.72 to 0.92; {\tt best\_dice} and {\tt best\_vi\_thr} collect, per case, the best Dice and the threshold that achieves it, over the 20 evaluated {\tt cases}.\footnote{We report results on only 20 cases, consistent with the artifact accompanying this paper, to keep testing times reasonable. Although these results are generally good, these 20 cases include some outliers. Therefore Dice scores for a much larger set of cases reported in~\cite{BelmonteCLM26}  are considerably higher.}

\subsection{Neuro-symbolic Whole Tumour Segmentation}\label{neuroMod}\label{neuroSym}

Program~\ref{lst:nnunet-body} trains a \nnunet\ model on the four modalities of the 50 training cases, with the expert annotation flattened to one whole-tumour class, and builds a predictor from it. All the configuration parameters of \nnunet\ are explicit in the program, so that the procedure can be inspected, audited and replicated. Most importantly, training and prediction are ordinary nodes of the plan: their results are stored in the cache, so running the specification again reuses the trained model and predictor instead of training the network again.

\begin{lstlisting}[style=imgql,float=tb,numbers=none,label={lst:nnunet-body},caption={Training \nnunet\ and using its prediction as reference (excerpt).}]
model     = nnunet.train(training_cases, work_root, modality_names,
                         nnunet_config, nnunet_folds, dataset_name,
                         nnunet_device, nnunet_trainer)
predictor = nnunet.make_predictor(model)
truth_of(i) = whole_tumour(intensity(nnunet.predict(predictor, modalities_of(i))))
\end{lstlisting}

With these definitions in place, the neuro-symbolic experiment requires a single change to Program~\ref{lst:sym}: the reference {\tt truth} in {\tt sweep\_of} becomes {\tt truth\_of(i)}, the prediction of the network, instead of the expert annotation. The threshold is thus chosen per case without any expert input. This mimicks the case in which one is dealing with new patient data for which no expert segmentation is available yet, which is the typical situation in a clinical setting. For comparison, the program also computes the Dice index, sensitivity and specificity of the predicted segmentation by the network itself against the expert annotation, and of the best symbolic segmentation against the segmentation predicted by the network.

\subsection{Results of the case study}\label{sec:cs-results}

\begin{table}[tb]
\centering
\caption{Whole-tumour segmentation, BraTS 2020 HGG cases as numbered in the dataset. \emph{Reference}: what the segmentation is scored against. \emph{Best per case}: the $\mathit{viThr}$ maximising Dice for each case. Sensitivity and specificity: means over cases (---: not computed).}
\label{tab:dice}
\footnotesize
\setlength{\tabcolsep}{3pt}
\begin{tabular}{@{}lllrrrrr@{}}
\toprule
& & & \multicolumn{3}{c}{\textbf{Dice}} & & \\ \cmidrule(lr){4-6}
\textbf{Segmentation} & \textbf{Reference} & \textbf{Cases} & \textbf{mean} & \textbf{median} & \textbf{s.d.} & \textbf{Sens.} & \textbf{Spec.} \\
\midrule
symbolic, $\mathit{viThr}=0.88$ & expert & 51--70 & 0.8276 & 0.8575 & 0.1285 & --- & --- \\
symbolic, best per case & expert & 51--70 & 0.8555 & 0.9074 & 0.1125 & --- & --- \\
\nnunet\ (trained 1--50) & expert & 51--70 & 0.8588 & 0.9152 & 0.1101 & 0.821 & 0.999 \\
symbolic, best per case & \nnunet & 51--70 & 0.8502 & 0.9259 & 0.1873 & 0.943 & 0.997 \\
\bottomrule
\end{tabular}
\end{table}

Table~\ref{tab:dice} collects the results of the various experiments. Against the expert annotation, tuning $\mathit{viThr}$ per case raises the mean Dice of the symbolic procedure from \csFixedDiceGt\ at the published fixed threshold to \csBestDiceGt. The best threshold varies from case to case (mean \csBestThrMeanGt, median \csBestThrMedianGt, range 0.82 to 0.92), which motivates the sweep. The \nnunet\ model reaches \csModelDice\ against the expert annotation on cases \csEvalCases. Used as the reference of the sweep, it yields a mean Dice of \csBestDiceNn\ between the symbolic segmentation and the network, which is close to the \csBestDiceGt\ obtained against the expert annotation. These results are very good: without any expert annotation, the symbolic procedure reproduces the segmentation of the network, and it explains it in terms that a domain expert can inspect. These percentages have to be interpreted keeping in mind  the relatively high intra-expert and inter-expert variability of 20$\pm$15\% and 28$\pm$12\%, respectively, for manual segmentations of brain tumour images~\cite{Mazzara2004}. Sensitivity and specificity of the best symbolic segmentation against the network are both close to 1, so few voxels are false positives or false negatives with respect to it. To keep the artifact easy to reproduce, this assessment uses 20 cases and does not compute all indexes (the dashes in Table~\ref{tab:dice}); the full study is left to follow-up work.

%

\section{Performance evaluation}\label{sec:perf}

We measure the engine on the threshold sweep of Section~\ref{OptSymGTV}:
\perfCases\ BraTS cases, 21 values of the very intense threshold each, and the
Dice index against the expert annotation for each value. The \nnunet\
experiment is left out: its cost is dominated by the GPU operations of the
network, not by the parallel scheduler. We study four aspects: how \vltwo\
compares with \vl\ on the same computation, how it uses the cores of a
workstation, the effect of element-wise fusion, and the savings due to the
content-addressed store when a program is repeated or interrupted.

\paragraph{Protocol.}
Measured on an Intel Core Ultra 9 285K (24 cores, 61\,GB), free-threaded CPython 3.14.7, SimpleITK 2.5.6, NumPy 2.3.2, Numba 0.67.0; 5 interleaved repetitions per configuration after one untimed warm-up run, each started on an otherwise idle machine.

Wall time, processor time and peak resident memory are measured by the
operating system (\texttt{wait4}). The engine's counters
(kernels executed, fusion cones, saturation) are taken from the summary it
prints when it exits. Each run starts on an empty store, except in the
experiments on the store itself.

\begin{table}[tb]
\centering
\caption{\vl\ against \vltwo, 40 cases $\times$ 21 thresholds. Wall and CPU summed over cases (median of 5 runs each); RSS: largest peak of any process. 1 of the 200 \vl\ runs failed with an internal error and was excluded.}
\label{tab:vl1}
\small
\setlength{\tabcolsep}{3pt}
\begin{tabular}{@{}lrrrr@{}}
\toprule
\textbf{Tool} & \textbf{Wall (s)} & \textbf{CPU (s)} & \textbf{RSS (GB)} & \textbf{Speedup} \\
\midrule
\vl, one process per case & 101.7 & 922 & 6.6 & 1.00 \\
\vltwo, one process per case & 102.7 & 1\,169 & 4.4 & 0.99 \\
\vltwo, one process for all cases & 45.2 & 924 & 6.4 & 2.25 \\
\bottomrule
\end{tabular}

\end{table}

\paragraph{\vl\ against \vltwo.}
\vl\ has neither loops nor sequences and cannot run the sweep as written. For
each case we therefore generate a \vl\ program in which the 21 thresholds are
unrolled, and a \vltwo\ program from the same template, so that the two tools
evaluate the same expressions. The two tools agree on all
\perfVlGoals\ Dice values to within $\perfVlMaxDiff$. Table~\ref{tab:vl1}
reports the cost. Case by case, the two tools take the same time
(\perfVlOneWall\,s for \vl\ and \perfVlTwoWall\,s for \vltwo), and \vl\ uses
\perfVlMemRatio\ times the peak memory. An interpreted implementation thus
matches a compiled one, because the execution time is spent in native imaging
kernels rather than in the interpreter. When all cases are given as a single
program using \texttt{for} loops, which is the intended use of \vltwo, \vltwo\ takes
\perfVlTwoAllWall\,s, \perfVlAllSpeedup\ times less than \vl, since the cases
are independent and are computed concurrently. \vl\ cannot run such a program,
because it does not release intermediate values once they are no longer
needed, and runs out of memory.

\begin{table}[tb]
\centering
\caption{Thread scaling of the sweep, median of 5 runs. \emph{Satur.}: achieved over requested concurrency; RSS: peak, in GB. Right: ITK pinned to one thread.}
\label{tab:scaling}
\footnotesize
\setlength{\tabcolsep}{4pt}
\begin{tabular}{@{}rrrrrrrr@{}}
\toprule
& \multicolumn{5}{c}{\textbf{ITK default}} & \multicolumn{2}{c}{\textbf{ITK = 1}} \\ \cmidrule(lr){2-6}\cmidrule(l){7-8}
\textbf{Workers} & \textbf{Wall (s)} & \textbf{Speedup} & \textbf{CPU/wall} & \textbf{Satur.} & \textbf{RSS} & \textbf{Wall (s)} & \textbf{Speedup} \\
\midrule
1 & 119.8 & 1.00 & 8.0 & 0.66 & 5.4 & 477.3 & 1.00 \\
2 & 73.8 & 1.62 & 12.8 & 0.78 & 3.8 & 243.5 & 1.96 \\
4 & 62.2 & 1.93 & 15.8 & 0.83 & 3.2 & 136.9 & 3.49 \\
8 & 56.9 & 2.11 & 17.6 & 0.90 & 5.5 & 86.6 & 5.51 \\
16 & 54.7 & 2.19 & 18.5 & 0.94 & 9.1 & 77.3 & 6.17 \\
24 & 54.6 & 2.19 & 18.5 & 0.91 & 12.0 & 92.5 & 5.16 \\
\bottomrule
\end{tabular}

\end{table}

\begin{figure}[tb]
\centering
\includegraphics[width=0.9\textwidth]{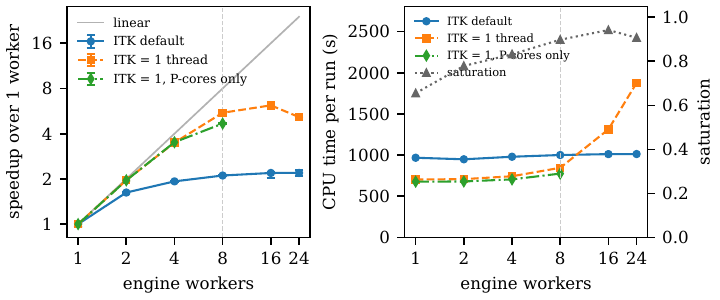}
\caption{Left: speedup over one engine worker (median and range), with linear
speedup for reference. Right: processor time per run, which stays flat while
added workers do the same work at the same cost, and saturation. Dashed: the
eight performance cores of the machine.}
\label{fig:scaling}
\end{figure}

\paragraph{Scaling with cores.}
Table~\ref{tab:scaling} and Fig.~\ref{fig:scaling} report the sweep for
increasing numbers of engine workers (the \texttt{--threads} option). Besides
wall time we report two quantities that explain a missing speedup. \emph{CPU
per wall}, processor time divided by wall time, is the average number of busy
cores. \emph{Saturation} is the number of kernels in execution (a fused cone counts as
one), sampled every 50\,ms and averaged over the run, divided by the number of
workers: if this number drops below 1, workers wait for work.

Workers and cores do not correspond one to one: the machine has 8 performance
and 16 slower efficiency cores, and each ITK filter starts its own threads. A
single worker keeps \perfCpuPerWallI\ cores busy, so the speedup at 24
workers, \perfSpeedupXXIV, is relative to a baseline that is already parallel.
In the right part of Table~\ref{tab:scaling} ITK is pinned to one thread, and
the number of workers equals the number of cores used. In that configuration
the speedup grows to \perfItkOneSpeedupVIII\ at 8 workers and
\perfItkOneSpeedupXVI\ at 16, and decreases at 24. With ITK unrestricted, the
fastest configuration (24 workers, \perfWallXXIV\,s) is faster than the best
pinned one (\perfItkOneBestWall\,s at \perfItkOneBestThreads\ workers), so
parallelism within a filter and parallelism between filters combine.

Beyond 8 workers, saturation stays at \perfSaturationXVI\ and the cores are
busy (CPU per wall \perfCpuPerWallXXIV\ at 24 workers), but wall time only
improves from \perfWallVIII\,s to \perfWallXXIV\,s. The workers are fed, so the
scheduler is not the bottleneck; we attribute the loss to the efficiency
cores, on which the additional workers run (Fig.~\ref{fig:scaling} also shows
the pinned configuration confined to the eight performance cores). Future work will be devoted to further optimising the
engine by investigating these bottlenecks.

\begin{table}[tb]
\begin{minipage}[t]{0.49\textwidth}
\centering
\caption{Fusion on and off, 24 workers, median (range) of 5 runs.}
\label{tab:fusion}
\scriptsize
\setlength{\tabcolsep}{3pt}
\begin{tabular}{@{}lrr@{}}
\toprule
\textbf{} & \textbf{Fusion on} & \textbf{Fusion off} \\
\midrule
Wall (s) & 52.1 {\tiny(51.7--56.0)} & 59.0 {\tiny(58.1--65.1)} \\
CPU (s) & 1\,007 & 1\,102 \\
Peak RSS (GB) & 9.0 & 8.0 \\
Peak live (GB) & 3.8 & 5.1 \\
Kernels & 26\,344 & 26\,344 \\
Cones & 5\,866 & 0 \\
Fused ops & 8\,801 & 0 \\
\bottomrule
\end{tabular}

\end{minipage}\hfill
\begin{minipage}[t]{0.48\textwidth}
\centering
\caption{Reuse through the store. \emph{Kernels}: primitive operations executed (an interrupted run prints none).}
\label{tab:store}
\scriptsize
\setlength{\tabcolsep}{3pt}
\begin{tabular}{@{}lrr@{}}
\toprule
\textbf{Run} & \textbf{Wall (s)} & \textbf{Kernels} \\
\midrule
\multicolumn{3}{@{}l}{\emph{Repeating a program}} \\
first, empty store & 52.5 {\tiny(51.9--55.7)} & 26\,344 \\
second, same store & 1.39 {\tiny(0.86--3.27)} & 101 \\
third, same store & 0.86 {\tiny(0.83--0.88)} & 9 \\
\multicolumn{3}{@{}l}{\emph{Interrupting and resuming}} \\
uninterrupted & 52.3 {\tiny(51.5--53.1)} & 26\,344 \\
stopped with Ctrl-C & 26.6 {\tiny(26.2--27.2)} & --- \\
restarted & 29.7 {\tiny(29.0--32.2)} & 15\,087 \\
\bottomrule
\end{tabular}

\end{minipage}
\end{table}

\paragraph{Element-wise fusion.}
Table~\ref{tab:fusion} compares the sweep with fusion (Section~\ref{tool})
enabled and disabled. Both configurations execute the same kernels; with
fusion, \perfFusionOps\ operations are grouped into cones, and disabling it
multiplies the median wall time by \perfFusionSpeedup. Fusion
also lowers the peak of live values from \perfFusionLiveOff\ to
\perfFusionLiveOn\,GB, because the intermediate images of a cone are never
stored, although the peak resident set does not decrease with it.

\paragraph{Reuse through the store.}
Table~\ref{tab:store} shows the effect of the store. Run again on the same
store, the sweep takes \perfWarmWall\,s instead of \perfColdWall\,s, and
executes at most \perfWarmKernelsMax\ of its \perfColdKernels\ kernels. A run
stopped with Ctrl-C half way and then restarted executes only
\perfResumeShare\% of the kernels again, and the two parts together take
\perfStoppedPlusResumedWall\,s, against \perfUninterruptedWall\,s for an
uninterrupted run.


\section{Related Work}\label{sec:related}

Manual segmentation is still the standard for {\em in vivo} images. However, this method is expensive and time-consuming, difficult to reproduce and possibly inaccurate due to human error. In areas where large, reliable datasets are available, machine learning and deep learning approaches have shown promising results in pattern recognition. Deep learning is based on the use of artificial neural networks, consisting of several layers, that can extract a hierarchy of features from raw input data. These methods often depend heavily on the availability of large training datasets. An interesting recent survey on such methods can be found in Akkus et al.~\cite{Akkus2017} and in work by Xia et al.~\cite{Ye25,Ze+25}. The latter work is based on self-supervised learning and semi-supervised medical image segmentation of CT scans. The work in~\cite{Ze+25} exploits the enlarged discrepancies at the feature level to train differential decoders and then learn from these differential features in an iterative way. 

Spatial model checking for the analysis of medical images is relatively recent and there are only a few articles exploring this field so far. In work by Sundstrom et al.~\cite{SGBM16}, spatio-temporal model checking techniques are used for the detection of tumours. The work is inspired by that of Grosu et al.~\cite{Gr+09} and aims at pursuing machine learning of the logical structure of image features. In contrast, the tool \vltwo focuses on human-intelligible logical descriptions that can be reused and extended. 
In earlier work by Grosu et al.~\cite{Gr+08}, a variant of spatial logic is proposed where spatial properties are expressed using quad trees. 
The combination of the latter logic with linear time signal temporal logic has led to the spatio-temporal logic SpaTeL~\cite{SPATEL}. A further variant combining spatial logic and signal temporal logic is the Signal Spatio-Temporal Logic (SSTL)~\cite{Ne+18}. There a  bounded surround operator is introduced that is inspired by the surround operator of an earlier version of the logic \slcs{.}

Some machine learning approaches require the generation of manual \emph{ground truth} labels, i.e. data sets in which segments of interest are indicated by experts manually in a standardised way. This is a complicated task not only because it is very laborious, but also because of the relatively high intra-expert and inter-expert variability of 20$\pm$15\% and 28$\pm$12\%, respectively, for manual segmentations of brain tumour images~\cite{Mazzara2004}. Nevertheless, much of the recent research in medical imaging and segmentation tasks in particular, has focused on these (probabilistic) learning algorithms (see for example~\cite{Menze2015,brats17preproc}). Approaches based on spatial model checking may be of help to improve the generation of manual ground truth labels in a more efficient, transparent and reproducible way. 

\section{Conclusions}\label{sec:conc}
We introduced \vltwo, a neuro-symbolic image analyser that combines neural network and logical primitives in an easy to use declarative language. We illustrated the development process, and how we can increase our confidence in AI generated code via differential testing and comparison with a reference implementation. We compared the new tool with its predecessor \vl, and the interpreted solution is on par with the compiled one in terms of performance. When the whole sweep is given to it as a single program, \vltwo is \perfVlAllSpeedup\ times faster than \vl\ running one program per case. The neuro-symbolic procedure, illustrated in~\cite{BCM25} is executed here with one, simple declarative program, giving an auditable and understandable procedure. On the twenty cases of Section~\ref{sec:test}, calibrating
one threshold against the network's prediction reaches a mean Dice of
\csBestDiceNn\ against that prediction and \csBestDiceGt\ against the expert
annotation, with each step of the procedure fully inspectable.

As future work, we plan to integrate formal verification in the development of \vltwo, in order to obtain formal guarantees about the correctness of the coordination and scheduling layer. Testing gives us confidence in the programs we run, while formal analysis of the coordination and scheduling model can ensure correctness on any run, even with millions of nodes involved. 

The nondeterminism of training remains an open problem: a network trained twice
on the same data need not give the same predictions, and hence the same
calibrated thresholds. Future work should study mechanisms to make models
``deterministic enough'', with results that are stable and depend only on a
pseudo-random seed, and the stability of \nnunet\ and other networks under
retraining in different random conditions. Moreover, we plan to validate the tool with experiments involving physicians, in order to evaluate usability and clarity of the tool.

%
%
%
\bibliographystyle{splncs04}
\bibliography{main}

\end{document}